# Spin-voltage-driven efficient terahertz spin currents from the magnetic Weyl semimetals Co$_2$MnGa and Co$_2$MnAl

*Genaro Bierhance[1,2], Anastasios Markou[3], Oliver Gueckstock[1,2], Reza Rouzegar[1,2], Yannic Behovits[1,2], Alexander Chekhov[1,2], Martin Wolf[1], Tom S. Seifert[1,2], Claudia Felser[3], Tobias Kampfrath[1,2,a]*

*Affiliations:*

[1]Department of Physical Chemistry, Fritz Haber Institute of the Max Planck Society, 14195 Berlin, Germany

[2]Freie Universität Berlin, Fachbereich Physik, Arnimallee 14, 14195 Berlin, Germany

[3]Max Planck Institute for Chemical Physics of Solids, Nöthnitzer Str. 40, 01187 Dresden, Germany

[a]Author to whom correspondence should be addressed: tobias.kampfrath@fu-berlin.de

*Abstract:* Magnetic Weyl semimetals are an emerging material class that combines magnetic order and a topologically non-trivial band structure. Here, we study ultrafast optically driven spin injection from thin films of the magnetic Weyl semimetals Co$_2$MnGa and Co$_2$MnAl into an adjacent Pt layer by means of terahertz emission spectroscopy. We find that (i) Co$_2$MnGa and Co$_2$MnAl are efficient terahertz spin-current generators reaching efficiencies of typical 3d-transition-metal ferromagnets such as Fe. (ii) The relaxation of the spin current provides an estimate of the electron-spin relaxation time of Co$_2$MnGa (165 fs) and Co$_2$MnAl (102 fs), which is comparable to Fe (92 fs). Both observations are consistent with a simple analytical model and highlight the large potential of magnetic Weyl semimetals as spin-current sources in terahertz spintronic devices. Finally, our results provide a strategy to identify magnetic materials that provide maximum spin current amplitudes for a given deposited optical energy density.

Magnetic Weyl semimetals (WSM) are a new class of materials with properties that are interesting from both a scientific and applied viewpoint [1-5]. Model representatives are the ferromagnetic Heusler compounds $Co_2MnGa$ (CMG) and $Co_2MnAl$ (CMA). CMG exhibits symmetry-protected, topological nodal-line band crossings and drumhead surface states [6] whereas CMA is considered a nodal-line WSM candidate [7, 8]. Their high spin polarization at the Fermi energy [3, 9, 10], the large anomalous Hall effect [7, 11] and anomalous Nernst effect [12, 13] make them promising candidates for efficient spin injection into an adjacent material.

It is highly interesting to push the speed of spin injection to femtosecond time scales and, thus, THz bandwidth by excitation with femtosecond laser pulses, as it was shown for other magnetically ordered materials previously [14, 15]. On one hand, the resulting THz spin transport (TST) can be used to generate spin torque [16, 17], to switch magnetic order [18] and to generate THz electromagnetic pulses for spectroscopy and photonic applications [19-27]. On the other hand, TST provides insight into the electron-spin relaxation time [28], which describes the time it takes until electron and spin degrees of freedom equilibrate with each other [18]. Recent studies on $Co_2MnSi$ and $Co_2FeAl$ have already shown the high potential of Co-based Heusler alloys in the context of spintronic terahertz emitters [29, 30].

Here, we use THz emission spectroscopy to characterize the photoinduced ultrafast spin-injection efficiency of CMG and CMA thin films into an adjacent Pt layer and the shape of the resulting spin-current pulses. A comparison with the standard ferromagnet Fe reveals a high efficiency of magnetic WSMs CMG and CMA as ultrafast spin-current generators. The relaxation time of the spin current pulse is dominated by the electron-spin equilibration, for which an upper limit of 165 fs and 102 fs is inferred for CMG and CMA, respectively. These values are 79% and 11% higher than for the Fe reference.

The principle of our THz emission experiments is illustrated in Fig. 1 [19, 31, 32]. The samples are sub||F|Pt stacks consisting of a ferromagnetic layer F and a Pt layer on an optically transparent substrate sub. The magnetization $M$ of layer F is saturated by an external magnetic field of the order of 100 mT. Magnetic hysteresis loops are shown in Fig. S1. The F|Pt thin-film stack is excited by a femtosecond pump pulse (duration ≈20 fs, center wavelength 800 nm, energy ≈2 nJ, repetition rate ≈80 MHz) from a Ti:sapphire laser oscillator. The excitation induces a transient spin voltage that launches a spin current with density $j_s$ from the ferromagnetic layer F into the Pt layer. Due to spin-orbit interaction, the spin current in Pt is converted into a transverse charge current perpendicular to the sample magnetization by the inverse spin Hall effect (ISHE), giving rise to the emission of a THz electromagnetic pulse (see Fig. 1).

The emitted THz pulse is measured by electro-optic sampling in ZnTe(110) or GaP(110) crystals (thickness 1 mm and 250 µm, respectively) [33]. The resulting signal $S(t)$ vs time $t$ is related to the THz electric field $E(t)$ directly behind the sample by a linear transfer function that accounts for the propagation of the THz pulse to the electro-optic crystal and the detection process. To connect $S(t)$ to

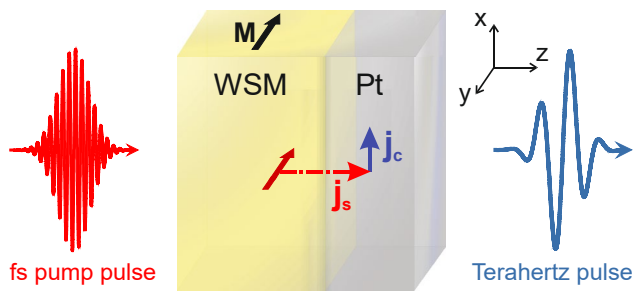

**FIG. 1. Laser-induced spin injection from a magnetic WSM.** The sample is a F|Pt stack where the ferromagnetic layer consists of the WSM CMG or CMA or the reference magnet Fe. Excitation by a femtosecond (fs) pump pulse launches a spin current with density $j_s$ from the F layer with magnetization $M$ into the Pt layer. Spin-orbit interaction converts $j_s$ into a transverse charge current $j_c$ that acts as a source of a THz electromagnetic pulse. The spin-to-charge-current conversion is assumed to be dominated by the inverse spin Hall effect in Pt.

the THz signal to the spin current, we note that $E(t)$ is in the frequency domain given by [19]

$$E(\omega) \propto Z(\omega)\gamma J_s(\omega). \qquad (1)$$

Here, $\omega/2\pi$ is the frequency, $Z(\omega)$ is the sample impedance, $\gamma$ is the spin Hall angle of Pt, and $J_s(\omega) = \int_0^d dz\, j_s(\omega, z)$ is the spin current density $j_s$ integrated over the Pt-layer thickness. Because the Pt layers of our samples are thicker than the relaxation length of $j_s$ ($\approx 1$ nm [34]), the spatial dependence of $j_s$ is similar in all studied samples. Equation (1) neglects a possible spin-to-charge current conversion in the F layer and its interfaces [35-37] because the spin Hall conductivity of Pt thin films is about 4-5, 2-3 and 4-5 times larger than the anomalous Hall conductivity of, respectively, Fe, CMA and thin film CMG [7, 11, 38, 39].

Note that the root-mean-square (rms) of $J_s(t)$ scales with the deposited pump energy density [inset of Fig. 2(a)], that is, $\mathrm{rms}J_s \propto A/d$, where $A$ is the pump absorptance of our sample, and $d$ is the thickness of the metal stack. Thus, $\mathrm{rms}J_s \, d/A$ provides a figure of merit (FOM) that characterizes the spin-injection efficiency of the F material into Pt. Following Eq. (1), we, therefore, compare [32]

$$\mathrm{FOM} = \frac{d}{A|Z|}\mathrm{rms}\,S. \qquad (2)$$

for F|Pt samples with F=CMG and CMA to a reference sample with F=Fe. In Eq. (2), we replaced the THz rms of the electric field $E$ behind the sample by the rms of the THz signal $S$, since $E$ and $S$ are

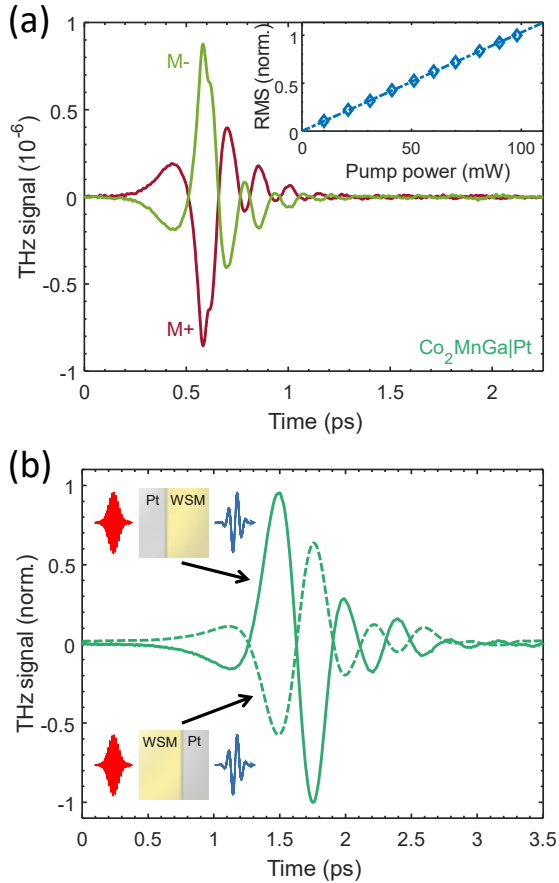

FIG. 2. **THz emission from a WSM|Pt stack.** (a) THz-emission signal $S(t, \pm M)$ from CMG|Pt for magnetizations $\pm M$ as obtained with a 250 μm thick GaP(110) electro-optic detector. Inset: Normalized THz-emission-signal amplitude (root-mean-square, rms) of the signal component odd in $M$ vs pump power. The dashed line is a linear fit through the origin. (b) THz emission signals $S(t)$ odd in $M$ [see Eq. (3)] from MgO||CMG|Pt and the same sample physically turned by 180° (Pt|CMG||MgO). The signals are obtained with a 1 mm thick ZnTe electro-optic detector and corrected for propagation effects (dispersion, absorption) through the MgO substrate [28]. Thus, the curves refer to CMG|Pt and Pt|CMG films with an MgO substrate on either side.

**Table 1. Sample parameters.** The columns show the sample, pump absorptance $A$, total metal-stack thickness $d$ and DC conductivity $G_{DC}^{4pp}/d$ averaged over $d$, where the DC sheet conductance $G_{DC}^{4pp}$ is determined with a four-point probe system. The absolute value $\overline{|Z|}$ of the sample sheet impedance (averaged over 1-5 THz), the DC conductivity $G(0)/d$ (averaged over $d$) and the Drude model damping rate $\Gamma_{eff}$ are obtained by THz transmission spectroscopy. For the sample structure, the numbers in brackets denote the layer thickness in nanometer.

| Sample | $A$ | $d$ (nm) | $\overline{|Z|}$ ($\Omega$) | $G_{DC}^{4pp}/d$ (MS/m) | $G(0)/d$ (MS/m) | $\Gamma_{eff}/2\pi$ (THz) |
|---|---|---|---|---|---|---|
| MgO\|\|Co$_2$MnGa(20)\|Pt(2)\|Si(3) | 0.61 | 25 | 39 | 0.71 | 0.60 | 19 |
| MgO\|\|Co$_2$MnAl(20)\|Pt(3)\|Si(3) | 0.63 | 26 | 50 | - | 0.35 | 25 |
| MgO\|\|Fe(3)\|Pt(3) | 0.54 | 6 | 43 | 2.13 | 2.07 | 19 |

related by an identical transfer function for all samples measured in our experiment. Finally, the sample impedance $Z(\omega)$ was approximated by $\overline{|Z|}$, i.e., $|Z(\omega)|$ averaged over 1-5 THz.

We study two-layer stacks Fe(3 nm)|Pt(3 nm), CMG(20 nm)|Pt(2 nm) and CMA(20 nm)|Pt(3 nm) on top of a MgO substrate (0.5 mm) [11]. The CMG|Pt and CMA|Pt films are additionally capped with Si(3 nm). The samples are grown by magnetron sputtering and structurally characterized by X-ray reflectivity. The sample impedance $Z(\omega)$ is determined by THz transmission spectroscopy [35], and its modulus is found to change by less than 2% over the interval 1-5 THz [see Figs. S2, S3]. This observation justifies the approximation $|Z(\omega)| = \overline{|Z|}$ introduced above. The sample absorptance $A$ is inferred by optical transmission and reflection measurements using the pump beam [35]. Table 1 compiles the values of $A$ and $\overline{|Z|}$ and other quantities of our samples.

Figure 2(a) displays typical THz signals $S(t, \boldsymbol{M})$ for opposite sample magnetizations $\pm \boldsymbol{M}$. The signal reverses almost completely with $\boldsymbol{M}$, showing that the THz signal is predominantly of magnetic origin. As we are only interested in effects odd in the sample magnetization, we focus on the signal

$$S(t) = \frac{S(t,\boldsymbol{M}) - S(t,-\boldsymbol{M})}{2} \tag{3}$$

in the following. The signal components even in $\boldsymbol{M}$ are about one order of magnitude smaller (see Fig. S4).

Figure 2(b) shows the THz emission signal $S(t)$ from MgO||CMG|Pt. When the sample is turned by 180°, resulting in Pt|CMG||MgO, the sign of $j_s$ and thus the THz signal reverses [35] but exhibits a larger amplitude. Note that the altered propagation of the pump pulse to the metal film and the THz pulse away from the metal film due to sample turning were corrected for by a suitable reference measurement [28]. We ascribe the different emission amplitudes from the measurements of MgO||CMG|Pt and Pt|CMG||MgO to the different amplitude of the pump field at the CMG/Pt interface. For MgO||CMG|Pt, the pump field at the CMG/Pt interface is significantly smaller than for Pt|CMG||MgO because the pump field is attenuated by 20 nm of CMG before reaching the interface.

Calculations of the pump propagation inside the metal stack show that the substrate-corrected amplitude ratio of the pump intensity at the CMG/Pt interface in the Pt|CMG sample and in the reversed CMG|Pt sample amounts to 1.34 (see Fig. S5), which is in reasonable agreement with the amplitude ratio of 1.59 of the two THz signals in [Fig. 2(b)]. We conclude that the emission from CMG|Pt has electric-dipole symmetry and is determined by the pump-field amplitude at the CMG/Pt interface [28].

Figure 3(a) shows scaled THz signals $S(t)d/A\overline{|Z|}$ from the three F|Pt samples, where F is the Fe reference and the magnetic WSMs CMG and CMA. The scaling factor $d/A\overline{|Z|}$ allows for a direct comparison in terms of spin-injection efficiency [see Eq. (2)].

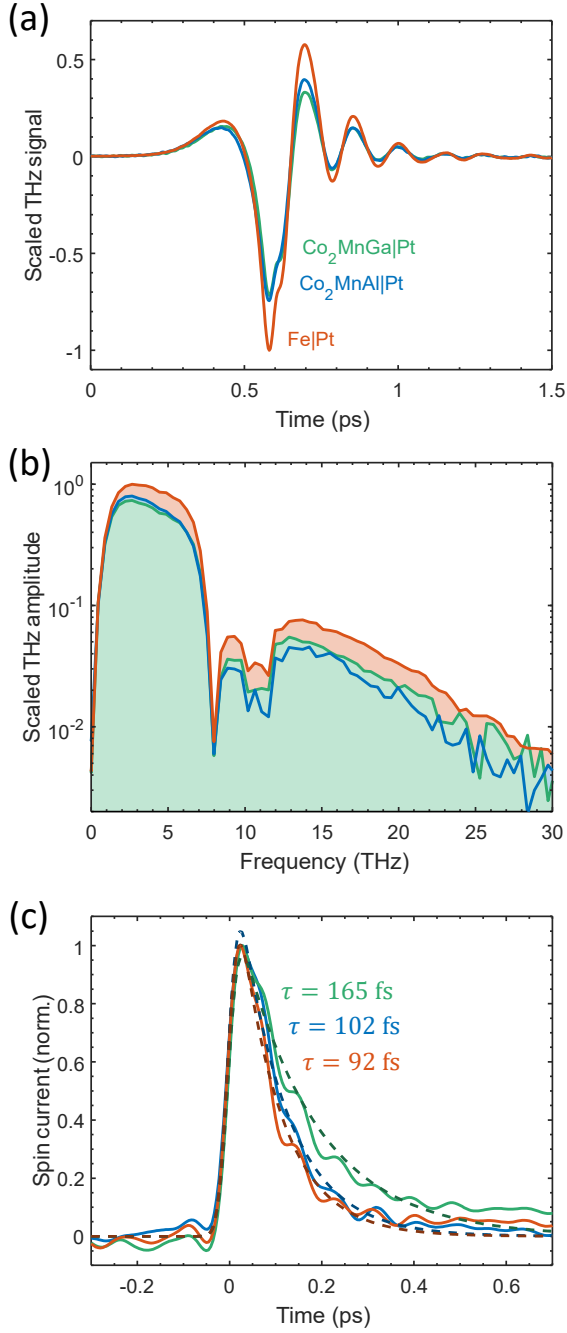

**FIG. 3. Comparison of THz spin transport in WSM|Pt vs Fe|Pt.** (a) Scaled THz emission signals $S(t)d/A\overline{|Z|}$ odd in $M$ from CMG|Pt, CMA|Pt and Fe|Pt as obtained with a 250 μm thick GaP electro-optic detector. Here, $A$, $\overline{|Z|}$ and $d$ are the pump absorptance, THz impedance and thickness of the sample. The scaled signal is a figure of merit in terms of spin-injection efficiency [see Eq. (2)]. (b) Corresponding frequency spectra for the waveforms in (a). (c) Normalized spin-current dynamics extracted from the THz-emission signals in (a). The dashed lines are single-sided monoexponential fits convoluted with a Gaussian (40 fs full width at half maximum) along with the obtained relaxation times $\tau$. In panels (a) and (b), signals are normalized to the peak signal from the Fe|Pt reference.

We find that the three signals exhibit very similar dynamics. The relative rms amplitudes of the three signals are, therefore, a direct measure of the spin-injection efficiency. Note that both CMG|Pt and CMA|Pt reach, respectively, 71% and 76% of the rms amplitude of Fe, which is in good agreement with the ratio of the peak values seen in Fig. 3(a). Thus, the two WSMs have a FOM characterizing spin-injection efficiency that can compete with that of Fe.

Figure 3(b) shows the amplitude spectra of the three THz emission signals. As expected from the time-domain signals [Fig. 3(a)], the spectra have very similar shape and comparable amplitude. We note that the zero at 8 THz and the feature at 12 THz are signatures of the electro-optic detection process in GaP [31, 33, 40].

To extract the dynamics $J_s(t)$ of the spin current, we measured the transfer function of our setup, used it to calculate back from the measured THz signal $S(t)$ to the THz field $E(t)$ and eventually employed Eq. (1) [41, 42]. The resulting currents $J_s(t)$ are shown in Fig. 3(c). As expected from the THz signals [Fig. 3(a)], the spin currents show approximately identical dynamics for all three samples: an instantaneous rise that is limited by our effective time resolution of about 40 fs [28], followed by a decay on a 100 fs time scale.

The extracted $J_s(t)$ can be well fit by an exponentially decaying Heaviside step function $\Theta(t)e^{-t/\tau}$ convoluted with a Gaussian with 40 fs full width at half maximum, which accounts for our effective time resolution. We obtain time constants of $\tau = 165$ fs, 102 fs and 92 fs for CMG|Pt, CMA|Pt and Fe|Pt, respectively. Previous work [28] has shown that the inverse time constant $\tau^{-1} = \tau_{ep}^{-1} + \tau_{es}^{-1}$ is a sum of the inverse time constants of electron-phonon ($\tau_{ep}^{-1}$) and electron-spin ($\tau_{es}^{-1}$) equilibration. In simple ferromagnetic metals such as $Co_{70}Fe_{30}$ or $Ni_{80}Fe_{20}$, $\tau_{ep}$ is typically substantially longer than $\tau_{es}$ [28]. Therefore, we expect that the time constant $\tau$ measured for CMG and CMA is a good estimate of $\tau_{es}$.

Note that $\tau_{es}^{-1}$ is a sum of terms due to electron spin flips and spin transport out of F. However, for F|Pt stacks with simple ferromagnetic metals F, transport was found to make a minor contribution [28]. Thus, $\tau$ is a lower bound to the time $\tau_{es}$ it takes electron and spin degrees of freedom to equilibrate with each other due to spin flips in F.

To summarize, our measurements show that (i) the efficiency of laser-induced spin injection of the magnetic WSMs CMG and CMA is almost as high (71% and 76%, respectively) as that of Fe [Fig. 3(a)]. (ii) The electron-spin relaxation time $\tau_{es}$ of CMG and CMA is only 79% and 11% longer than for Fe [Fig. 3(c)]. Both results appear surprising at first glance because CMG and CMA are rather different materials than Fe as they possess semimetal-like cone-shaped electronic bands near the Fermi energy. However, in addition to these topologically protected bands, there are "trivial" bands that contribute a significant electronic density of states at the Fermi energy.

To discuss the results (i) and (ii) in a more quantitative manner, we make use of the simple analytic spin-flip and transport model of Ref. [28]. In the framework of this model, observation (ii) points to similar values of the product $P_{sf}D^\uparrow D^\downarrow/\chi_m$, which quantifies the contribution of electronic spin flips to the ultrafast decay of the spin voltage of the ferromagnetic layer F (see Ref. [28]). Here, $P_{sf}$ scales with the spin-flip matrix element, $D^\uparrow$ and $D^\downarrow$ denote the density of states of spin-up and spin-down electrons at the Fermi level of F, and $\chi_m$ is the magnetic susceptibility. Indeed, the values of $D^\uparrow$ and $D^\downarrow$ are quite comparable for Fe ($D^\uparrow \approx 0.8$ and $D^\downarrow \approx 0.2$/(eV atom) [43], CMG [$D^\uparrow \approx 1.5$ and $D^\downarrow \approx 0.2$/(eV atom)] [44] and CMA [$D^\uparrow \approx 1.5$ and $D^\downarrow \approx 0.3$/(eV atom)] [45].

Because the measured $\tau_{es}$ of Fe and CMG are similar, we conclude that the ratio $P_{sf}/\chi_m$ of spin-flip matrix element and magnetic susceptibility of CMG can compete with that of Fe. This result is plausible because Fe and CMG exhibit similar values of $D^\uparrow$ and $D^\downarrow$ and roughly comparable Curie temperatures [11], which, in the framework of the Stoner model, suggest comparable values of $\chi_m$ [46]. On the other hand, the anomalous Hall effect of CMG [47] and CMA [7] and the anomalous Nernst effect of CMG [48] and, thus, spin-orbit coupling and the value of $P_{sf}$ in CMG and CMA are sizeable.

Regarding result (i), the spin current amplitude directly after excitation by the pump can be shown to roughly scale with $M_0'(T_0)\Delta T_{e0}T_{tr}^\uparrow/\tau_{es}P_{sf}D^\downarrow$ (see Ref. [28]). The value of $\tau_{es}P_{sf}D^\downarrow$ is similar for Fe and CMG, as discussed for observation (ii) above. We also expect similar values for the gradient $M_0'(T_0)$ of the F magnetization $M_0$ vs equilibrium temperature $T_0$: While the saturation magnetization of Fe is about twice larger than that of CMG, this ratio is compensated by the lower Curie temperature of CMG [11, 49] (700 K vs 1000 K), resulting in a larger gradient $M_0'(T_0)$ at $T_0 = 300$ K. The pump-induced initial increase $\Delta T_{e0}$ of the electron temperature is quite similar in the two materials because of their comparable electronic density of states [10, 43, 44]. Consequently, the observed similar spin-current magnitudes [Fig. 3(a)] suggest that the transmission coefficient $T_{tr}^\uparrow$ of spin-up electrons through the F/Pt interface has similar magnitude both for F=Fe and CMG. To summarize, our observations (i) and (ii) are also plausible from the viewpoint of the simple analytical model of Ref. [28].

In conclusion, we have measured ultrafast spin current generation by femtosecond laser pulses in CMG|Pt and CMA|Pt stacks. We find that (i) for a given excitation density, the spin current amplitudes are comparable to Fe|Pt stacks, thereby identifying CMG and CMA as efficient ultrafast light-driven spin-current injectors or, in other words, efficient sources of transient spin voltages. (ii) The relaxation time of the spin current is somewhat longer than for Fe and provides an estimate of the electron-spin relaxation time of CMG and CMA due to spin flips. Both results are consistent with a simple analytical model [28]. Therefore, our consideration of the spin current amplitude can be used as a general strategy to identify F-material candidates that deliver maximum spin current amplitudes for a given deposited optical energy density for highly efficient THz spintronic devices.

See the Supplementary Material for further experimental details.

## Acknowledgments

The authors thank Martin Jourdan (Johannes-Gutenberg-Universität Mainz, Germany) for providing the Fe|Pt sample. They acknowledge funding by the European Union H2020 program through the FET projects SKYTOP/Grant No. 824123 and the German Research Foundation through the collaborative research center SFB TRR 227 "Ultrafast spin dynamics" (projects A05 and B02).

## Data availability

The data that support the findings of this study are available from the corresponding author upon reasonable request.

## Conflict of interest

The authors have no conflicts to disclose.

**Supplementary Material**

**Sample preparation.** Heteroepitaxial growth on MgO [11], for layer thickness see Table 1 (main text).

CMG and CMA films were grown using magnetron sputtering in a BESTEC ultrahigh vacuum system with a base pressure less than $2 \times 10^{-9}$ mbar and a process gas (Ar 5 N, i.e. purity 99.999%) pressure of $3 \times 10^{-3}$ mbar. The target to substrate distance was fixed at 20 cm and the substrates were rotated during deposition to ensure homogeneous growth. CMA was grown from stoichiometric compound $Co_2MnAl$ (5.08 cm) source using 40 W DC power at 550 °C followed by *in situ* post annealing for 24 min at the same temperature. CMG was grown from Co (5.08 cm) Mn (5.08 cm) and MnGa (5.08 cm) sources in confocal geometry, using 34 W, 6W and 20 W DC power, respectively, at 600 °C. The film was then annealed *in situ* under UHV at 600 °C for 25 min. Both films were capped with Pt and Si films, at room temperature, from Pt (5.08 cm) and Si (5.08 cm) using 25 W DC and 60 W RF power, respectively. The growth rates and the film thicknesses were determined by using x-ray reflectivity measurements.

The fabrication of the Fe|Pt sample on MgO is detailed in Ref. [19].

**Sample characterization.** The optical pump absorptance (800 nm center wavelength) for the investigated samples (Table 1) is determined by optical transmittance and reflectance measurements using a power meter [35].

THz conductivities are obtained from THz transmission measurements under dry air atmosphere. We use a THz spintronic emitter (W|CoFeB|Pt) as the THz radiation source [19]. The emitter is excited by 800 nm pump pulses (more details about the laser system are specified in the next section). Residual pump radiation is blocked with a high resistivity Si wafer behind the emitter. THz signals are obtained by electro-optic sampling using 250 µm thick GaP. The bare MgO substrate serves as reference. The detailed procedure to obtain the frequency-resolved conductivities based on a Drude fit model from is described in Ref. [34]. Conductivity and impedance results are shown in Figs. S2 and S3.

The saturation magnetization is determined by the magneto-optic Faraday effect. To this end, we use pulses from a Ti:sapphire laser oscillator (center wavelength of 800 nm, pulse duration 30 fs, repetition rate 80 MHz, pulse energy 0.9 nJ). The polarization rotation of optical pulses after transmission through the samples (Faraday effect) is measured by a combination of a Wollaston prism and two balanced photodiodes as a function of an external in-plane magnetic field from an electromagnet [50]. To obtain a significant projection of the in-plane magnetization onto the beam propagation direction, we choose an optical angle of incidence of 45°. The resulting hysteresis loops are shown in Fig. S1 for CMG|Pt and CMA|Pt. They show that the magnetization in CMG and CMA saturates for external fields larger than 30 mT and 60 mT, respectively, indicating that the experiments in this work are performed under a saturating external magnetic field.

**Terahertz emission spectrometer.** THz emission measurements are performed in dry-air atmosphere [Figs. 2(a), 2(b), 3(a)-(c), S4(a)-(c)]. The samples are excited by femtosecond pump pulses (duration ≈20 fs, center wavelength 800 nm, energy ≈2 nJ, repetition rate ≈80 MHz) from a Ti:sapphire laser oscillator. THz signals are obtained by electro-optic sampling using 1 mm thick ZnTe(110) [Fig. 2(b)] and 250 µm thick GaP(110) [Figs. 2(a), 3(a)-(c), S4(a)-(c)]. For the comparison in Figs. 2(a) and 2(b) the displayed signals were shifted on the time axis such that maxima (or minima) of the odd in the magnetization *M* signals intersect.

**Correction for substrate propagation.** To determine the effect of a pump intensity gradient inside a thin film heterostructure on the THz emission amplitude obtained from MgO||CMG|Pt (sub configuration) and the 180°-turned Pt|CMG||MgO (cap configuration), we remove the substrate influence on pump incoupling and THz outcoupling while the sample is physically turned as follows. A correction function is obtained from THz emission spectra of the sufficiently thin reference (ref) sample MgO||CoFe|Pt (sub) and 180°-turned Pt|CoFe||MgO (cap) sample as $S_{\mathrm{ref}}^{\mathrm{cap}}(\omega)/S_{\mathrm{ref}}^{\mathrm{sub}}(\omega)$. With a total film thickness of 7 nm, we expect an approximately constant pump intensity along the $z$ axis (Fig. 1) for the reference. Above 8 THz the cap spectra are dominated by the noise floor due to the increasing THz absorption of MgO [51]. The result of the correction procedure is shown in Fig. 2(b) and further discussed in the main text.

**Pump-intensity distribution.** The pump electric field inside the metallic stacks is calculated by a generalized $4 \times 4$ matrix formalism [52, 53] using Matlab codes provided in Ref. [53]. As a cross check, we also use numerical calculations in Comsol, leading to consistent results. The refractive indices of MgO, CMG and Pt at a wavelength of 800 nm are taken from Refs. [54-56].

The resulting pump intensity profiles are shown in Fig. S5 for MgO||CMG|Pt and the 180°-turned Pt|CMG||MgO. We obtain a pump intensity ratio of 0.84 at the Pt/CMG and CMG/Pt interfaces. Note that this value still contains differences in the pump propagation through the sample, which differs for MgO||CMG|Pt vs Pt|CMG||MgO. The experimental results displayed in Fig. 2(b) are already corrected for THz propagation out of and pump propagation into the sample. To remove the influence of pump incoupling, we multiply the calculated intensity ratio with $t_{A,MgO}^2 n_{MgO}^2$, where $n_{MgO}$ is the refractive index of the substrate, and $t_{A,MgO}$ is the Fresnel transmission coefficient of the air/substrate interface. The resulting value of 1.34 is an estimate of the pump-gradient-related changes in the emitted THz-field amplitude for Pt|CMG vs CMG|Pt and in reasonable agreement with the experimentally obtained value of 1.59.

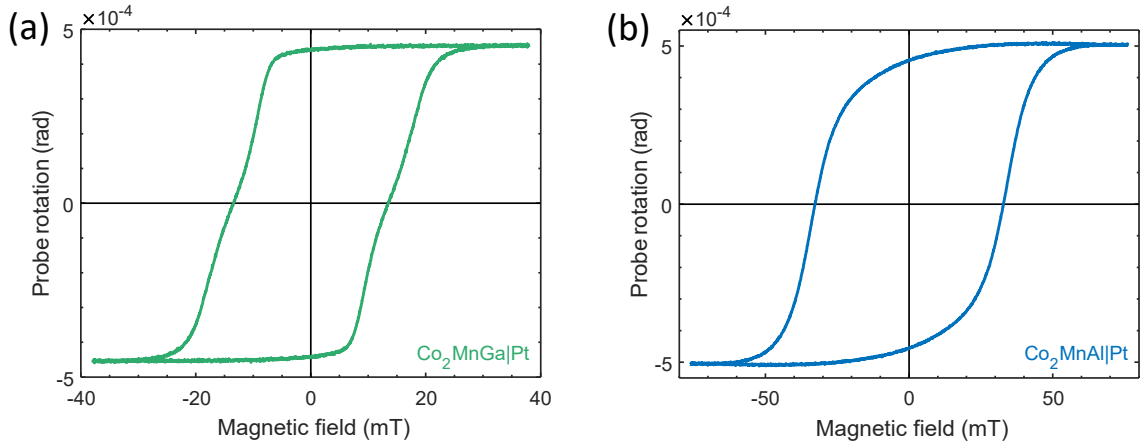

**FIG. S1.** Magnetic hysteresis loops measured by Faraday rotation. (a) Probe polarization rotation upon transmission through CMG|Pt vs applied magnetic field. (b) Same as (a), but for CMA|Pt.

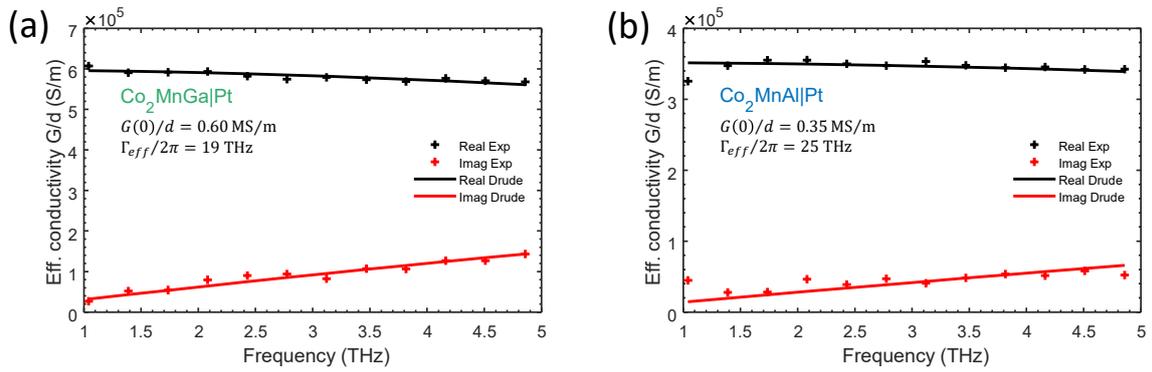

**FIG. S2.** (a) Mean conductivity $G/d$ of the CMG|Pt stack along with a fit based on the Drude model [34]. The best-fit parameters $G(0)/d$ (DC conductivity) and $\Gamma_{\text{eff}}$ (effective current relaxation rate) are given in the figure. (b) Same as (a), but for CMA|Pt.

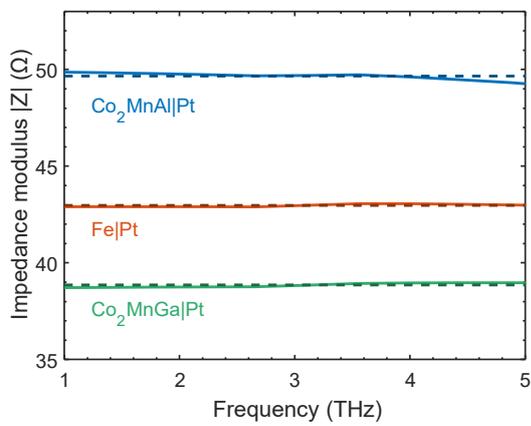

**FIG. S3.** Modulus of the sample impedance as determined from the Drude-model fit results of Fig. S2. Dashed lines correspond to averages over 1-5 THz.

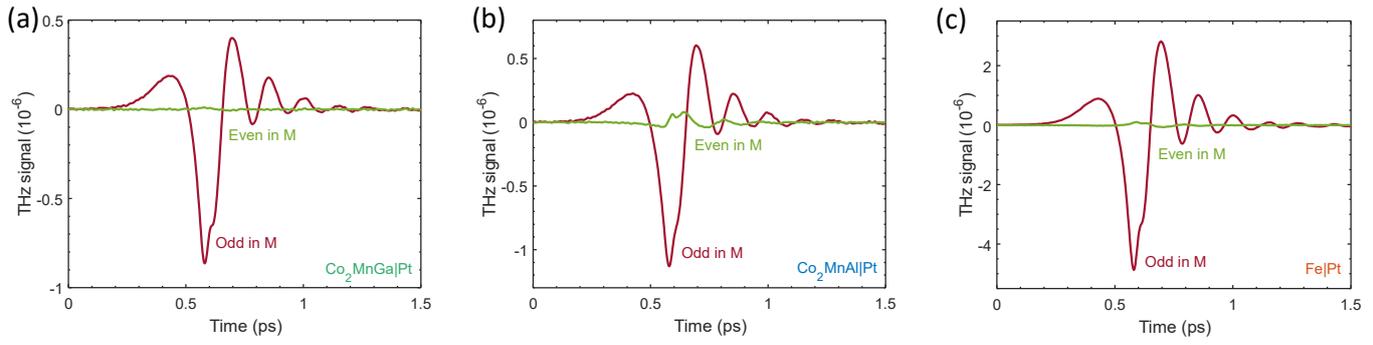

**FIG. S4.** (a) THz-emission signal components $[S(t,\boldsymbol{M}) \pm S(t,-\boldsymbol{M})]/2$ even/odd in the magnetization $\boldsymbol{M}$ of the CMG|Pt sample. (b), (c) Same as (a), but for CMA|Pt and Fe|Pt.

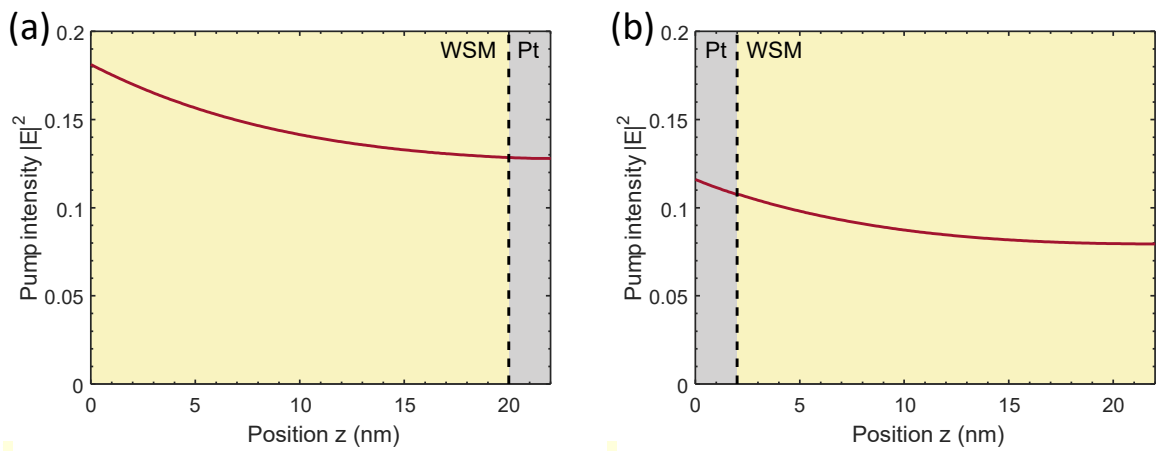

**FIG. S5.** (a) Pump intensity vs $z$ (see Fig. 1) in MgO∥CMG|Pt and (b) the 180°-turned sample Pt|CMG∥MgO.